\pdfoutput=1

\documentclass{PoS}
\usepackage{amsmath}
\usepackage{verbatim}

\title{Fractional electric charge and quark confinement}

\ShortTitle{Fractional electric charge and quark confinement}

\author{\speaker{Sam R. Edwards}\\
       Institut f\"ur Kernphysik, Technische Universit\"at Darmstadt, D-64289 Darmstadt, Germany\\
        E-mail: \email{edwards@crunch.ikp.physik.tu-darmstadt.de}}

\author{Andr\'e Sternbeck\\
        Institut f\"ur Theoretische Physik, Universit\"at Regensburg, D-93040 Regensburg, Germany\\
        E-mail: \email{andre.sternbeck@physik.uni-regensburg.de}}

\author{Lorenz von Smekal\\
        Institut f\"ur Kernphysik, Technische Universit\"at Darmstadt, D-64289 Darmstadt, Germany\\
        E-mail: \email{lorenz.smekal@physik.tu-darmstadt.de}}

\abstract{
Owing to their fractional electric charges, quarks are blind to transformations that combine a color center phase with an appropriate electromagnetic one. Such transformations are part of a global $Z_6$-like center symmetry of the Standard Model that is lost when quantum chromodynamics (QCD) is treated as an isolated theory. This symmetry and the corresponding topological defects may be relevant to non-perturbative phenomena such as quark confinement, much like center symmetry and ordinary center vortices are in pure SU($N$) gauge theories. Here we report on our investigations of an analogous symmetry in a 2-color model with dynamical Wilson quarks carrying half-integer electric charge. 
}

\FullConference{The XXIX International Symposium on Lattice Field Theory - Lattice 2011\\
July 10-16, 2011\\
Squaw Valley, Lake Tahoe, California}

\begin{document}

\section{Introduction}
In spite of tremendous research efforts, the phase diagram of QCD matter at non-zero temperature and chemical potential is not well understood. There remains much to learn about the exact nature of the transition of hadronic matter into a deconfined plasma in the hot and dense conditions of heavy ion collisions and the early universe. 

For the finite temperature deconfinement transition of infinitely
heavy quarks, at least, the importance of center symmetry has been
established \cite{Greensite:2003bk}. The deconfinement transition of
pure SU($N$) gauge theory in $d+1$ dimensions mirrors the $Z_N$3
symmetry breaking of a $d$-dimensional $Z_N$ spin model
\cite{Svetitsky:1982gs}, with spacelike center vortices playing the
role of spin interfaces. Their percolation at low temperatures leads
to confinement by disordering the Polyakov loop, while their dynamical
suppression at high temperatures allows for an ordered, deconfined
phase \cite{DeForcrand:2001dp,deForcrand:2001nd}. The connection with
spin models is especially striking for the 
second order phase transitions of SU(2) and SU(3) in 2+1 
dimensions \cite{Edwards:2009qw,Strodthoff:2010dz}. Here the
self-duality of the corresponding 2-dimensional 
spin models is manifested in the universal behavior of center vortex
and electric flux free energies, which are reflections of one another
about criticality \cite{vonSmekal:2010xz}. 

The role of center symmetry is, however, obscured by the inclusion of
dynamical quarks in the fundamental representation. Center symmetry is
explicitly broken and the finite temperature transition of QCD becomes
a smooth crossover at zero chemical potential
\cite{Borsanyi:2010bp,Bazavov:2010sb}.   

This neglects the electric charge of quarks, which is commonly
expected to only require small perturbative corrections. Note, though,
that the inclusion of the quarks' fractional electric charge leads to
a global $Z_6$-symmetry that combines the centers of the color and
electroweak gauge groups. The physical realization of this symmetry
could have non-trivial implications for confinement and the phase
structure of the Standard Model. In these proceedings, we report on
our ongoing study of such a symmetry in 2-color QCD with half-integer
electrically charged quarks.  

\section{Hidden symmetry}
We will give only a brief overview of the global center symmetry of the Standard Model and its analog in our toy lattice model. The interested reader is directed to our proceedings from last year's lattice conference \cite{Edwards:2010ew}, as well as \cite{Baez:2009dj} for a review of the group structure and \cite{Bakker:2005ph} for a different lattice realization of the symmetry.

Since quarks carry fractional electric charges $Q=\frac{2}{3}e$ or $-\frac{1}{3}e$, the color and electromagnetic phases in the pair of combined transformations
\begin{equation}
(e^{i2\pi/3},e^{i2\pi Q/e}),\;(e^{-i2\pi/3},e^{-i2\pi Q/e})\in \text{SU(3)}\times\text{U(1)}_{em} 
\end{equation}
cancel precisely. What's more, they act trivially on all other particles in the Standard Model, which are blind to the center of SU(3) and carry integer electric chage. Electric charge is related to hypercharge $Y$ and the third component of weak isospin $t_3$ by $Q/e=t_3+Y/2$, with $e^{i2\pi t_3} \equiv -1 \in$ SU(2) and $Y$ quantized in units of $1/3$. The symmetry is therefore generated by 
\begin{equation}
(e^{i2\pi /3},-1,e^{i\pi Y})\in \text{SU(3)}\times \text{SU(2)}\times \text{U(1)}_Y,
\end{equation}
which gives six elements including the identity.

A global symmetry brings with it the possibility of a phase transition characterized by spontaneous symmetry breaking. In this case, one expects the transition to be driven by topological defects (vortices) that carry both color and electromagnetic flux. That is, center vortices with an additional electromagnetic Dirac string. Could this be relevant to (de)confinement?

\section{A toy 2-color world}
Our starting point is 2-color QCD plus electromagnetism with 2 flavors of Wilson fermions in $3 + 1$ dimensions. By including `up' and `down' quarks with fractional charge $\pm \frac{1}{2}e$ relative to the U(1)$_{em}$ gauge action, we obtain a model with a global Z$_{2}$ symmetry. The lattice action is 
\begin{equation}
S = -\sum_{\text{plaq.}}\left( \frac{\beta _{col}}{2}\text{Re Tr }U_p +\beta _{em} \cos \theta_p\right) +  S_{f,W},
\end{equation}
where $S_{f,W}$ is the usual Wilson fermion action with the distinction that parallel transporters for quarks are  products of an SU(2) color matrix and a U(1)$_{em}$ phase, of the form
\begin{equation}
\label{eqn:transporters}
 U_\mu(x) e^{i\theta_{\mu}(x)/2},\;\; U_\mu (x)\in\text{SU(2)},\;\theta_{\mu}(x)\in (-2\pi,2\pi].
\end{equation}
The SU(2) plaquettes $U_p$ and U(1)$_{em}$ plaquette angles $\theta_p$ are formed from $U_\mu$ and $\theta_\mu$ in the usual way. In this model, 'fractional charge' means that the parallel transporters for quarks contain \emph{half} the U(1)$_{em}$ angle relative to the $\theta_\mu$ that appear in the plaquette angle $\theta_p$. That is, an $e^{i\theta_\mu/2}=-1$ electromagnetic link for quarks appears as an $e^{i\theta_\mu}=+1$ link in the gauge action. An important point is that the 'size' of the compact U(1)$_{em}$ is determined by the quark, which is the smallest quantum of electric charge. The range of $\theta_{\mu}(x)$ is chosen such that we integrate over all possible electromagnetic transporters for the quarks, amounting to a double counting for the U(1)$_{em}$ gauge action and integer charged particles. This is consistent with the premise that our compact U(1)$_{em}$ is the result of symmetry breaking in a SU(3)$
\rightarrow$SU(2)$\times$U(1)$_{em}$ unified theory (see e.g. \cite{Preskill:1984gd}). 

It is clear from Eq. \eqref{eqn:transporters} that a combined color center phase $-1\in$ SU(2) and electromagnetic phase $e^{i\theta_\mu/2}=-1$ acts identically on the quarks. The model therefore retains a combined color and electromagnetic $Z_2$ center symmetry, despite the introduction of dynamical quarks. This is the analog of the hidden $Z_6$ symmetry of the Standard Model.

\section{Simulations}
\subsection{Z$_2$ disorder}
Our preliminary results in \cite{Edwards:2010ew} indicated that the usual ordering of color links by dynamical quarks is negated in our model by the inclusion of fractional electric charge. This is understood by analogy with spin systems. In QCD alone, terms in a loop expansion of the fermion determinant that wind around the temporal direction favor the center sector in which the traced Polyakov loop $P_{col}=1$. They break center symmetry and lead to ordering in much the same way as spins coupled to an external magnetic field. The inclusion of fractional electric charge in our model bestows quark loops with an additional U(1)$_{em}$ phase which may undo this effect. Consider, for example, the hopping expansion of our Wilson fermion action for $N_t=4$ time slices to fourth order in the hopping parameter $\kappa$,
\begin{equation}
\label{eqn:hopping}
S_{f,\text{eff}} = -16\kappa^4 \left( \sum_{\text{plaq.}} \cos{\frac{\theta_p}{2}}\cdot \text{Re Tr }U_p  + 8\sum_{\vec{x}} \cos{P_{\theta/2}} \cdot {\text{Re }{P_{col}}  } \right) +\dots ,
\end{equation}
If the U(1)$_{em}$ Polyakov loop angle for quarks $P_{\theta/2}(\vec{x})=\sum_{t=0}^{N_t-1}\theta_t(t,\vec{x})/2$ is disordered, then the $P_{col}=1$ sector is no longer favored and SU(2) center symmetry is \emph{dynamically} restored. Since the parallel transporters for quarks possess a Z$_{2}$ degree of freedom that the U(1)$_{em}$ gauge action is blind to, this is possible even in the Coulomb phase for integer electric charges. The effect is analogous to placing an Ising model in a fluctuating magnetic field, or to the Peccei-Quinn mechanism \cite{Peccei:1977}, in which the coupling of the CP violating term in QCD to an axion field allows for the dynamical restoration of CP symmetry.

Indeed, we found in \cite{Edwards:2010ew} that the SU(2) Polyakov loop in our toy model was indistinguishable from the quenched SU(2) result at values of the hopping parameter that would otherwise cause a significant amount of ordering in standard 2-color QCD ($\kappa=0.15,\; 0.175$). For random, `hot' starts, where the U(1)$_{em}$ links are totally disordered, this was true both for the confined $\beta_{em} \lesssim 1.01$ and Coulomb phases $\beta_{em} \gtrsim 1.01$ of the U(1)$_{em}$ gauge action. Since the U(1)$_{em}$ gauge action is blind to Z$_{2}$ disorder as seen by quark, i.e. $e^{i\theta_\mu /2}=\pm  1$, it is unable to remove it.

On the other hand, Z$_{2}$ disorder was not able to be generated deep in the Coulomb phase for `cold', ordered starts ($U_\mu=1$, $e^{i\theta_\mu/2}=1$). Here we found the usual behavior for 2-color Wilson fermions, which we attibuted to the ergodicity limitations of our Hybrid Monte Carlo (HMC) algorithm. In the Coulomb phase, the many simultaneous fluctuations in the U(1)$_{em}$ angles $\theta_{\mu}(x)$ proposed by the HMC update are strongly suppressed. This prohibits the transformation of, e.g., an $e^{i\theta_\mu /2}=1$ link to an $e^{i\theta_\mu /2}=-1$ link, even when they are equivalent in the U(1)$_{em}$ gauge action.

\begin{figure}[htbp]
  \begin{minipage}[t]{0.48\columnwidth}
    \centering
		\includegraphics[width=\columnwidth]{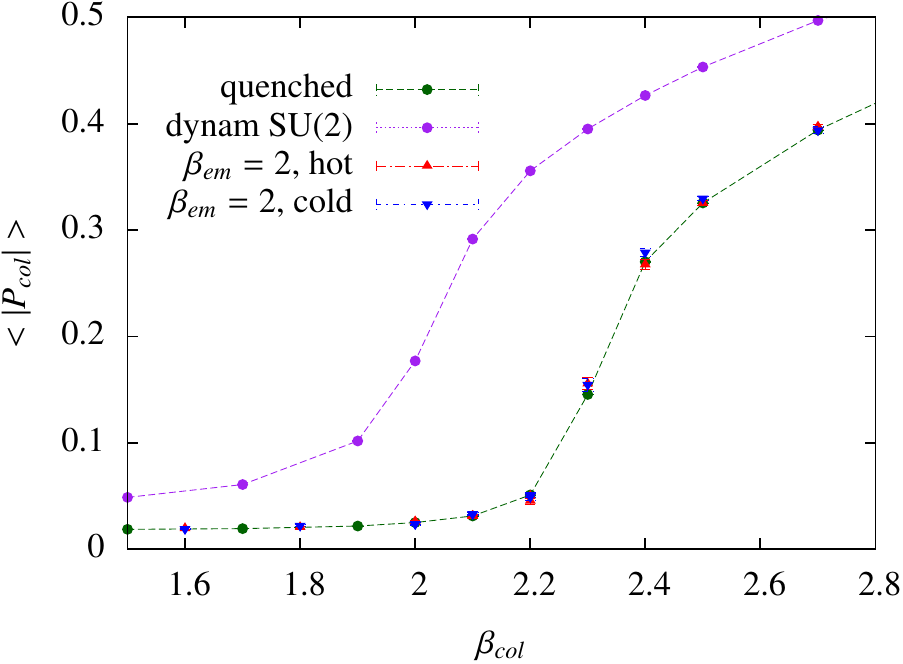}
	\caption{Volume average of the traced SU(2) Polyakov loop on $8^3\times 4$ lattices with $\kappa = 0.15$ and $\beta_{em}=2$. The `quenched' and `dynam SU(2)' results are from 2-color QCD without electromagnetism. `Hot' and `cold' starts are in the prescence of quarks with fractional electric charge. They are consistent with the quenched SU(2) result, demonstrating that SU(2) center symmetry is dynamically restored.}
	\label{fig:plot-su2poly}
  \end{minipage}
  \hspace{0.45 cm}
  \begin{minipage}[t]{0.48\columnwidth}
    \centering
		\includegraphics[width=\columnwidth]{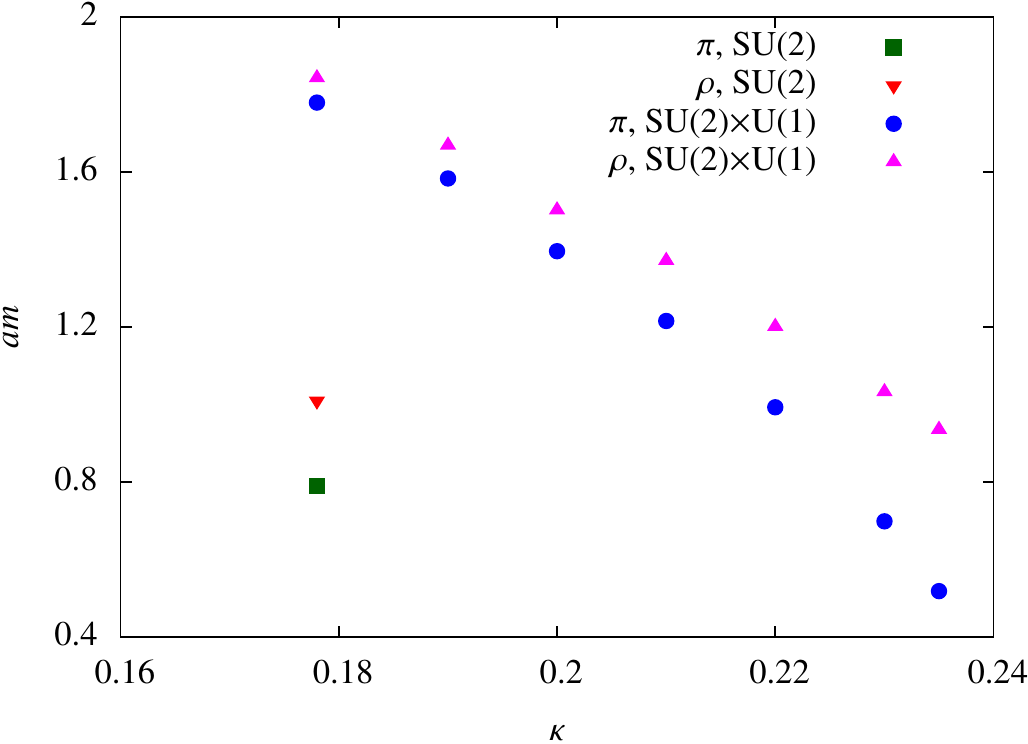}
	\caption{Meson masses on $8^3\times 16$ at $\beta_{col}=1.7$ and $\beta_{em}=2$. `SU(2)' results are from standard 2-color QCD. Inclusion of a compact U(1)$_{col}$ and fractional electric charge dramatically shifts the mass scale.  Much larger values of $\kappa$ are required in order to obtain an equivalent ratio of $\rho$ and $\pi$ masses at a given $\beta_{col}$. }
	\label{fig:masses}
  \end{minipage}
\end{figure}

We have since ameliorated this problem by including additional local updates. The first of which is a simultaneous transformation of SU(2) and U(1)$_{em}$ links, $U_{\mu}\rightarrow -U_{\mu},; e^{i\theta_\mu /2}\rightarrow -e^{i\theta_\mu /2}$. The only part of the action that is affected by this combined Z$_2$ update is the SU(2) Wilson plaquette term. The U(1)$_{em}$ gauge action is blind to such phases by default, and the quark determinant is also blind because this transformation belongs to our Z$_2$ symmetry. It introduces a thin color-electromagnetic vortex around the transformed link. Between HMC trajectories we propose such an update on a random link and accept/reject using a Metropolis check for the SU(2) gauge action, repeating many times. This is efficient when $\beta_{col}$ is small, but the acceptance rate is exponentially suppressed as the average value for the SU(2) plaquette increases with $\beta_{col}$.  It becomes necessary to also include updates that flip only a U(1)$_{em}$ link, $e^{i\theta_\mu /2}\rightarrow -e^{i\theta_\mu /2}$. The fermion action requires an expensive recalculation, but the acceptance rate is much improved for regions of parameter where $\beta_{col}$ is large and $\kappa$ is moderate. Given sufficient thermalization, `hot' and `cold' starts show complete agreement when such local updates are included. This is demonstrated in Fig. \ref{fig:plot-su2poly}, where we plot the traced SU(2) Polyakov loop in the U(1)$_{em}$ Coulomb phase at $\beta_{em}=2$ using the same  $\kappa=0.15$ as in \cite{Edwards:2010ew}. It is indistinguishable from the pure gauge result, showing the dynamical restoration of SU(2) center symmetry.

\newpage
\subsection{Mass scales}
The values for the hopping parameter $\kappa$ used in our test simulations were based on the 2-color results of Skullerud et al. \cite{Skullerud:2003yc}. Given the large impact that fractional electric charge has on the SU(2) Polyakov loop, however, it is not reasonable to expect the mass scale to remain unchanged and for these $\kappa$ values to remain appropriate. As such, we performed zero temperature measurements of meson masses on $8^3\times 16$ lattices at $\kappa=0.178$, $\beta_{col} = 1.7$, and $\beta_{em} = 2$ to directly compare with  \cite{Skullerud:2003yc}. Masses were extracted from all-to-all 2-point correlation functions with pointlike sources, using dilution in the spin and time indices in accordance with \cite{Foley:2005ac}.

The inclusion of fractional electric charge has a dramatic effect, even though we are in the Coulomb phase for the U(1)$_{em}$ gauge action. At these parameters, the $\pi$ and $\rho$ masses in lattice units are roughly twice those of the standard 2-color result (Fig. \ref{fig:masses}). This stems from two factors. Firstly, the inclusion of U(1)$_{em}$ links in the parallel transporters increases the disorder in the 2-point correlation functions. In particular, the Z$_2$ disorder in the phase of the U(1)$_{em}$ links for quarks leads to a Z$_2$ electromagnetic flux string between constituent quarks in addition to the usual color flux string. What's more, since the explicit breaking of SU(2) center symmetry by quarks is suppressed by the introduction of fractional electric charge, the color links are less ordered than in standard 2-color QCD. This is evident if one calculates meson masses using only the SU(2) links from our SU(2)$\times$U(1)$_{em}$ configurations. These masses, corresponding to  electrically neutral sources, are consistent with those calculated on \emph{quenched} SU(2) configurations at the same parameters. 

\begin{figure}[htbp]
  \begin{minipage}[t]{0.48\columnwidth}
    \centering
		\includegraphics[width=\columnwidth]{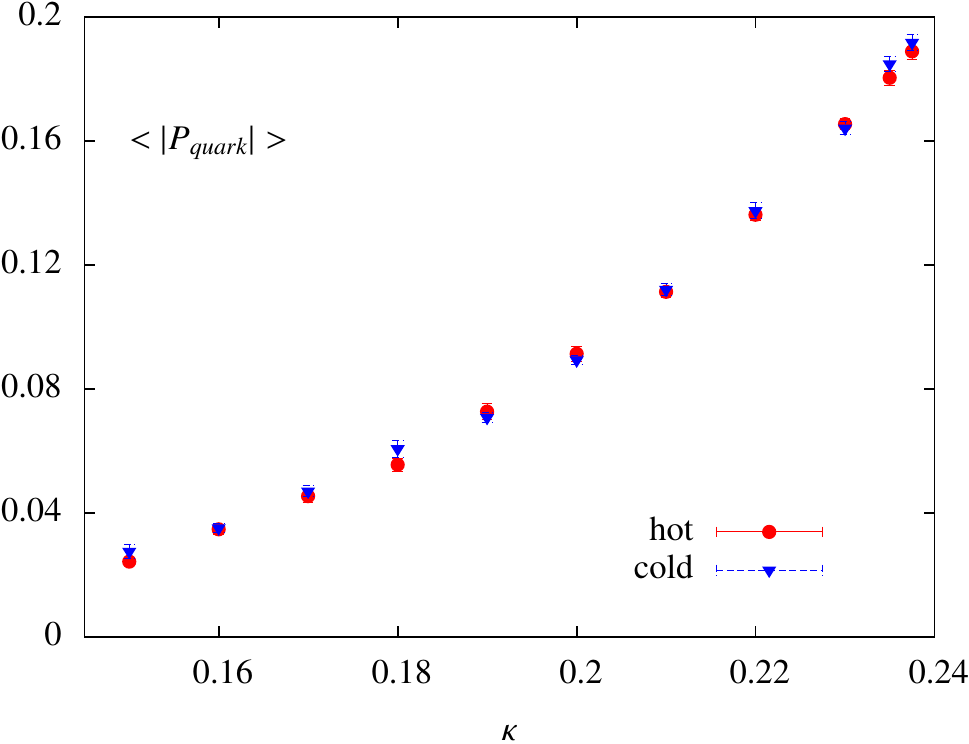}
	\caption{Volume averaged SU(2)$\times$U(1)$_{em}$ Polyakov loop corresponding to a static quark, $P_{quark} (\vec{x})=  \cos{P_{\theta /2}(\vec{x})} \cdot \text{Re } P_{col}(\vec{x})$,  on an $8^3\times 4$ lattice with $\beta_{col} = \beta_{em}=2$. SU(2) and U(1)$_{em}$ links order with respect to each other as the quark mass is decreased. }
	\label{fig:plot-su2u1order}
  \end{minipage}
  \hspace{0.45 cm}
  \begin{minipage}[t]{0.48\columnwidth}
    \centering
		\includegraphics[width=\columnwidth]{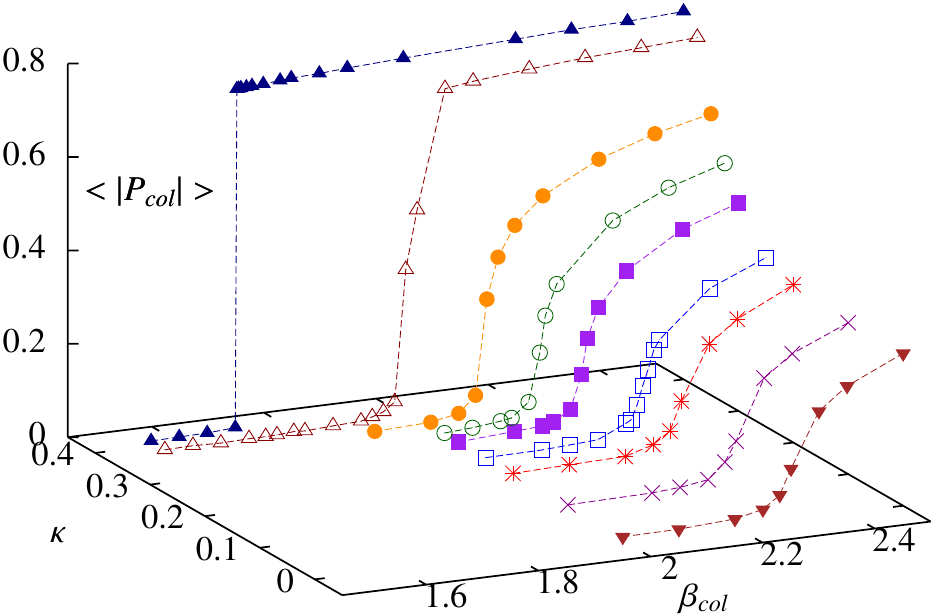}
	\caption{SU(2) Polyakov loop on $16^3 \times 4$ lattices, using an $\mathcal{O}(\kappa^4)$ hopping effective action at $\beta_{em}=\infty$. The location of the transition is shifted with increasing $\kappa$.}
	\label{fig:hopping}
  \end{minipage}
\end{figure}

This begs the question: has the inclusion of fractional electric charge with respect to our U(1)$_{em}$ gauge action effectively removed the quarks from the functional integral? Such a conclusion would be reasonable if the Z$_2$ U(1)$_{em}$ phases for the quarks were truly random. But quarks loops induce a coupling between the SU(2) and U(1)$_{em}$ links. Consider the plaquette-plaquette and Polyakov loop-Polyakov loop terms from our effective action Eq. \eqref{eqn:hopping}. These are minimized if the products are unity. That is, if the SU(2) and U(1)$_{em}$ links order with respect to one another. This ordering is evident in the real part of the trace of the combined SU(2)$\times$U(1)$_{em}$ Polyakov loop that corresponds to a quark line
\begin{equation}
 P_{quark} (\vec{x})=  \cos{P_{\theta /2}(\vec{x})} \cdot \text{Re } P_{col}(\vec{x}),
\end{equation} 
 which increases as $\kappa$ is increased, i.e. as the quark mass decreases (see Fig. \ref{fig:plot-su2u1order}). 

We have explored the phase diagram with the $\mathcal{O}(\kappa^4)$ effective action Eq. \eqref{eqn:hopping}, which allows us to cheaply check our intuition about the coupling of SU(2) and U(1)$_{em}$ links without having to worry about the chiral limit for $\kappa$. In Fig. (\ref{fig:hopping}) we show results for the SU(2) Polyakov loop with the U(1)$_{em}$ phases $e^{i\theta_\mu(\vec{x}) /2}$ restricted to $\pm 1$, which amounts to $\beta_{em}=\infty$ in the U(1)$_{em}$ gauge action. As $\kappa$ increases, the disorder-order transition of the SU(2) Polyakov loop moves to smaller values of $\beta_{col}$ and sharpens dramatically. Note that in the combined limit $\kappa,\; \beta_{em} \rightarrow \infty$, the plaquette-plaquette coupling in Eq. \eqref{eqn:hopping} forces the SU(2) plaquettes to take the values $\pm 1$. For very large $\kappa$ the transition line should therefore terminate with the first order bulk transition of Z$_2$ gauge theory at $\beta_{col} \sim 0.44$ \cite{Wegner:1984qt}. As the SU(2) plaquette is driven to unity for large $\beta_{col}$, U(1)$_{em}$ plaquettes corresponding to quark loops, $\cos {\frac{\theta_p}{2}}$, receive an effective coupling that suppresses Z$_2$ disorder in their phases. The relevant U(1)$_{em}$ Polyakov loop $\cos{P_{\theta/2}}$ remains disordered at small values of $\kappa$, but transitions to unity for large $\beta_{col}$ and $\kappa$ (not shown).  

The coupling SU(2) and U(1)$_{em}$ links by quarks therefore leads to results distinct from the pure gauge theory, but the interesting physics lies at much larger values of $\kappa$ than for standard 2-color QCD. In particular, the mass scale has been radically shifted by the inclusion of fractional electric charge. A ratio of $\rho$ and $\pi$ masses of $\sim 1.28$, which is obtained for $\kappa=0.178$ at $\beta_{col}=1.7$ in standard 2-color QCD, requires $\kappa \sim 0.223$ in our SU(2)$\times$U(1)$_{em}$ theory at $\beta_{em}=2$. The chiral limit is correspondingly pushed from $\kappa_c \sim 0.185$ to $\kappa_c \sim 0.241$ (see Fig. \ref{fig:masses}).

\newpage
\vspace{-.5 cm}

\section{Discussion}
\vspace{-.1 cm}
Given the fractional electric charge of quarks and the existence of a global symmetry that relates the centers of the color and electroweak gauge groups, it may be misleading to study non-perturbative phenomena in QCD alone. In our toy 2-color model, the coupling of Wilson quarks with half-integer electric charge relative to a compact U(1)$_{em}$ has a dramatic effect on the color sector. The SU(2) Polyakov loop has an order-disorder transition that is consistent with the spontaneous center symmetry breaking transition of the pure gauge theory at values of the hopping parameter where center symmetry is clearly \emph{explicitly} broken in standard 2-color QCD. 

The model brings with it additional technical problems, however. Hybrid Monte Carlo trajectories must be supplemented with additional local updates to ensure ergodicity, and the mass scale is shifted to much larger values of $\kappa$. The determination of a complete phase diagram will require considerable computational effort. As such, it is worthwhile to study fractional electric charge in even simpler models, such as our $\mathcal{O}(\kappa^4)$ hopping expansion model. Numerical simulations in an analogous gauge-Higgs theory are also ongoing \cite{Greensite:2012}.

\vspace{.3 cm}

\noindent\textbf{Acknowledgements:} This work was supported by the
Helmholtz International Center for FAIR within the LOEWE program of
the State of Hesse, the Helmholtz Association Grant VH-NG-332, and the
European Commission, FP7-PEOPLE-2009-RG No.~249203. Simulations were performed on the LOEWE-CSC High Performance Cluster at Goethe-Universit\"at Frankfurt am Main.


\begin{thebibliography}{99}	
\setlength{\itemsep}{-3pt}

\bibitem{Greensite:2003bk}
J.~Greensite, Prog.\ Part.\ Nucl.\ Phys.\ {\bf 51} (2003) 1.

\bibitem{Svetitsky:1982gs}
  B.~Svetitsky and L.~G.~Yaffe,
  Nucl.\ Phys.\  B {\bf 210} (1982) 423.

\bibitem{DeForcrand:2001dp}
  Ph.~de Forcrand and L.~von Smekal,
  Nucl.\ Phys.\ Proc.\ Suppl.\  {\bf 106} (2002) 619.

\bibitem{deForcrand:2001nd}
  P.~de Forcrand and L.~von Smekal,
  Phys.\ Rev.\  D {\bf 66}, 011504 (2002)
  [arXiv:hep-lat/0107018].

\bibitem{Edwards:2009qw}
  S.~Edwards and L.~von Smekal,
  Phys.\ Lett.\  B {\bf 681} (2009) 484.

\bibitem{Strodthoff:2010dz}
  N.~Strodthoff, S.~R.~Edwards and L.~von Smekal,
  PoS {\bf LATTICE2010}, 288 (2010)
  [arXiv:1012.0723 [hep-lat]].
  
\bibitem{vonSmekal:2010xz}
  L.~von Smekal, S.~R.~Edwards and N.~Strodthoff,
  PoS {\bf LATTICE2010}, 292 (2010)
  [arXiv:1012.0408 [hep-lat]].

\bibitem{Borsanyi:2010bp} 
  S.~Borsanyi {\it et al.}  [Wuppertal-Budapest Collaboration],
  JHEP {\bf 1009}, 073 (2010)
  [arXiv:1005.3508 [hep-lat]].
  
\bibitem{Bazavov:2010sb} 
  A.~Bazavov {\it et al.}  [HotQCD Collaboration],
  J.\ Phys.\ Conf.\ Ser.\  {\bf 230}, 012014 (2010)
  [arXiv:1005.1131 [hep-lat]].
 
 
\bibitem{Edwards:2010ew}
  S.~R.~Edwards, A.~Sternbeck and L.~von Smekal,
  PoS {\bf LATTICE2010}, 275 (2010)
  [arXiv:1012.0768 [hep-lat]].

\bibitem{Baez:2009dj}
  J.~C.~Baez and J.~Huerta,
  Bull.\ Amer.\ Math.\ Soc. {\bf 47} (2010) 483.

\bibitem{Bakker:2005ph}
  B.~L.~G.~Bakker, A.~I.~Veselov and M.~A.~Zubkov,
  Phys.\ Lett.\  B {\bf 620} (2005) 156 .

\bibitem{Preskill:1984gd}
  J.~Preskill,
  Ann.\ Rev.\ Nucl.\ Part.\ Sci.\  {\bf 34}, 461 (1984).
  
\bibitem{Peccei:1977}
  R.~D.~Peccei and H.~R.~Quinn,
  Phys.\ Rev.\ Lett.\  {\bf 38} (1977) 1440;   Phys.\ Rev.\  D {\bf 16} (1977) 1791.
  

    
\bibitem{Skullerud:2003yc}
  J.~I.~Skullerud, S.~Ejiri, S.~Hands and L.~Scorzato,
  Prog.\ Theor.\ Phys.\ Suppl.\  {\bf 153} (2004) 60.

\bibitem{Foley:2005ac}
  J.~Foley, K.~Jimmy Juge, A.~O'Cais, M.~Peardon, S.~M.~Ryan and J.~I.~Skullerud,
  Comput.\ Phys.\ Commun.\  {\bf 172}, 145 (2005)
  [arXiv:hep-lat/0505023].

\bibitem{Brown:1990ev}
  F.~R.~Brown {\it et al.},
  Phys.\ Rev.\ Lett.\  {\bf 65} (1990) 2491.


\bibitem{Wegner:1984qt}
  F.~J.~Wegner,
  J.\ Math.\ Phys.\  {\bf 12}, 2259 (1971).

\bibitem{Greensite:2012}
  J.~Greensite, K.~Langfeld, S.~Edwards, and L.~von Smekal,
  work in progress.

\end{thebibliography}
\end{document}